\documentclass[aps,prb,twocolumn,superscriptaddress,showpacs]{revtex4}
\usepackage{color}
\usepackage{graphicx}

\begin{document}


\title{Temperature Dependent Magnetic Anisotropy in (Ga,Mn)As Layers}


\author{M. Sawicki}
 \email{mikes@ifpan.edu.pl}
\affiliation{Institute of Physics, Polish Academy of Sciences, al.
Lotnik\'ow 32/46, 02-668 Warszawa, Poland}
\author{F. Matsukura}
\affiliation{Institute of Physics, Polish Academy of Sciences, al.
Lotnik\'ow 32/46, 02-668 Warszawa, Poland}
\affiliation{2Laboratory for Nanoelectronics and Spintronics,
Research Institute of Electrical Communication, Tohoku University,
Katahira 2-1-1, Aoba-ku, Sendai 980-8577,~Japan~and \\ERATO
Semiconductor Spintronics Project, Japan Science and Technology
Agency, Japan }
\author{A. Idziaszek}
\affiliation{Institute of Physics, Polish Academy of Sciences, al.
Lotnik\'ow 32/46, 02-668 Warszawa, Poland}
\author{T. Dietl}
\affiliation{Institute of Physics, Polish Academy of Sciences, al.
Lotnik\'ow 32/46, 02-668 Warszawa, Poland} \affiliation{ ERATO
Semiconductor Spintronics Project, al. Lotnik\'ow 32/46, PL-02668
Warszawa, Poland \\and Institute of Theoretical Physics, Warsaw
University, PL-00681 Warszawa, Poland}
\author{\\G.M. Schott}
\affiliation{Physikalisches Institut (EP3), Universit\"at
W\"urzburg, Am Hubland, D-97074 W\"urzburg, Germany}
\author{C. Ruester}
\affiliation{Physikalisches Institut (EP3), Universit\"at
W\"urzburg, Am Hubland, D-97074 W\"urzburg, Germany}
\author{C. Gould}
\affiliation{Physikalisches Institut (EP3), Universit\"at
W\"urzburg, Am Hubland, D-97074 W\"urzburg, Germany}
\author{G. Karczewski}
\affiliation{Institute of Physics, Polish Academy of Sciences, al.
Lotnik\'ow 32/46, 02-668 Warszawa,
Poland}\affiliation{Physikalisches Institut (EP3), Universit\"at
W\"urzburg, Am Hubland, D-97074 W\"urzburg, Germany}
\author{G. Schmidt}
\affiliation{Physikalisches Institut (EP3), Universit\"at
W\"urzburg, Am Hubland, D-97074 W\"urzburg, Germany}
\author{L.W. Molenkamp}
\affiliation{Physikalisches Institut (EP3), Universit\"at
W\"urzburg, Am Hubland, D-97074 W\"urzburg, Germany}

\date{\today}

\begin{abstract}
It is demonstrated by SQUID measurements that (Ga,Mn)As films can
exhibit perpendicular easy axis at low temperatures, even under
compressive strain, provided that the hole concentration is
sufficiently low.  In such films, the easy axis assumes a
standard in-plane orientation when the temperature is raised
towards the Curie temperature or the hole concentration is increased
by low temperature annealing. These findings are shown to
corroborate quantitatively the predictions of the mean-field Zener
model for ferromagnetic semiconductors. The in-plane
anisotropy is also examined, and possible mechanisms accounting for its
character and magnitude are discussed.
\end{abstract}

\pacs{75.50.Pp, 75.30.Gw, 75.70.-i}

\maketitle

\section{Introduction}

The discovery of carrier-mediated ferromagnetism in (III,Mn)V
dilute magnetic semiconductors (DMS) grown by molecular beam
epitaxy (MBE) has made it possible to combine complementary
properties of semiconductor quantum structures and ferromagnetic
systems in single devices, paving the way for the development of
functional semiconductor spintronics.\cite{Spintronics} Therefore,
the understanding of magnetic anisotropy in these systems and the
demonstration of methods for its control is timely and important.
It has been known, since the pioneering works of Munekata {\it et
al.}\cite{Mune93} and Ohno {\it et al.},\cite{Ohno96,Shen97} that
ferromagnetic (In,Mn)As and (Ga,Mn)As films are characterized by a
substantial magnetic anisotropy.
Shape anisotropy effects can explain neither the direction nor the
large magnitude of the observed anisotropy field $\mu_0 H_A$ in
these dilute magnetic materials.

It has been found by studies of the anomalous Hall effect
\cite{Ohno96,Shen97} and ferromagnetic resonance \cite{Liu03} that
the direction of the easy axis is rather controlled by epitaxial
strain in these systems. Generally, for layers under tensile
biaxial strain [like (Ga,Mn)As on an (In,Ga)As buffer]
perpendicular-to-plane magnetic easy axis has been observed
(perpendicular magnetic anisotropy, PMA). In contrast, the layers
under compressive biaxial strain [as canonical (Ga,Mn)As on a GaAs
substrate] have been found to develop in-plane magnetic easy axis
(in-plane magnetic anisotropy, IMA). At first glance this
sensitivity to strain appears surprising, as the Mn ions are in
the orbital singlet state $^6A_1$.\cite{Szczy99} For such a case
the orbital momentum $L = 0$, so that effects stemming from the
spin-orbit coupling are expected to be rather weak and, indeed,
electron paramagnetic resonance studies of Mn in GaAs have led to
relevant spin Hamiltonian parameters by two orders of magnitude
too small to explain the values of $\mu_0 H_A$.\cite{Fedo02}

In the system in question, however, the ferromagnetic Mn-Mn
exchange interaction is mediated by the band holes, whose
Kohn-Luttinger amplitudes are primarily built up of anion p
orbitals. Furthermore, in semiconductors, in contrast to metals,
the Fermi energy is usually smaller than the atomic spin-orbit
energy. Hence, as noted by some of the present authors and
co-workers,\cite{Diet97,Diet00} the confinement or strain-induced
anisotropy of the valence band can result in a sizable anisotropy
of spin properties. Indeed, the quantitative calculation within
the mean-field Zener model,\cite{Diet00,Diet01a,Abol01} in which
the valence band is represented by the 6$\times$6 Luttinger
Hamiltonian, explains the experimental values of $H_A$ in
(Ga,Mn)As with an accuracy better than a factor of
two.\cite{Diet01a,Diet01b} Moreover, by combining theories of
magnetic anisotropy\cite{Diet00,Diet01a} and of magnetic
stiffness,\cite{Koni01} it has been possible to describe the width
of stripe domains in (Ga,Mn)As PMA films.\cite{Diet01b} However,
the theories in question\cite{Diet01a,Abol01} contain a number of
predictions that call for a detail experimental verification.

In this paper we present magnetic anisotropy studies carried out
by direct magnetization measurements in a dedicated
superconducting quantum interference device (SQUID) magnetometer.
Our results show that films of (001) (Ga,Mn)As on GaAs with
appropriately low values of hole density $p$ do exhibit PMA and
that in some of them a clear temperature-induced reorientation of
the easy axis from [001] to $\langle 100\rangle$ occurs. This
peculiar behavior is actually quite universal and has been
observed by some of us and co-workers in the case of (Al,Ga,Mn)As
(Ref.~\onlinecite{Taka02}) and of (Ga,Mn)As obtained under
different growth conditions.\cite{Sawi03} Importantly, the effect
vanishes in samples with higher hole concentrations, which we
obtain by low temperature annealing. By a quantitative comparison
of these findings to results of theoretical computations, we
demonstrate that the temperature-induced reorientation of the easy
axis from the perpendicular (PMA) to the in-plane (IMA)
orientation corroborates the theoretical expectations referred to
above.\cite{Diet01a} At the same time our magnetization data
confirm a peculiar character of the {\em in-plane} anisotropy,
which can be inferred from previous transport\cite{Kats98} and
magnetoptical\cite{Hrab02,Welp03} measurements. We discuss
possible reasons and consequences of this specific
temperature-dependent in-plane magnetic anisotropy of the
(Ga,Mn)As/GaAs material system.

\section{Samples and experimental}

The (Ga,Mn)As films with a thickness of 400~nm were deposited by
MBE onto (001) GaAs substrates at 220$^o$C under an As$_4$/Ga beam
equivalent pressure ratio of 5, as described
previously.\cite{Scho01} High resolution x-ray diffraction shows
good crystal quality (Ga,Mn)As with the rocking curve widths
comparable to those of the GaAs substrate and pronounced
finite-thickness fringes, indicating flat interfaces and surfaces.
The layers are pseudomorphic with respect to the GaAs substrate,
as verified by reciprocal space maps around the asymmetric [115]
reflection. We present results for samples with Mn concentration
$x = 5.3$\% and $3$\%, and biaxial strain $\varepsilon_{xx} =
-0.27$\% and $-0.16$\%, respectively, as established by x-ray
diffraction measurements.\cite{Scho01} In order to trace the
evolution of magnetic anisotropy with the hole concentration, the
$x =5.3$\% sample was divided into six pieces, five of them  being
annealed at 165$^o$C in air for different times between 28 and 200
h, which led to an increase of $T_{\mbox{\tiny C}}$ up to a factor
of two. This sensitivity to annealing reflects the presence of
self-compensation by mobile donor defects, which are neutralized
at the surface.\cite{Edmo04}

Magnetization measurements are carried out in the SQUID
magnetometer down to 5~K.
By design, the signal detection axis is aligned with the direction
of the magnetic field ${\bm H}$ (vertically). As a result the
system is sensitive only to the vertical component of the
magnetization vector. Special care is put forward to screen the
sample from external magnetic fields and to keep the parasite
remanent fields generated by the magnet at the lowest possible
level. The Meissner effect of a pure lead sample confirms that
even after an excursion to the field of 1~kOe, the remanent field
remains below 100~mOe. Such a low value is essential for
successful magnetic remanence studies or for measurements at
temperatures close to $T_{\mbox{\tiny C}}$, where the coercivity
of (Ga,Mn)As drops into a sub-Oersted regime. For studies of
in-plane anisotropy, (Ga,Mn)As layers were initially shaped by
chemical etching into circles of approximately 5~mm in diameter.
However, detailed studies shows that this is an unnecessary
precaution, since even rectangular 2:1 shapes have the same
properties as circled samples.

\section{Origin of magnetic anisotropy in ferromagnetic zinc-blende DMS}

The magnetic dipolar anisotropy, or shape anisotropy, is mediated
by dipolar interaction. Since it is long range, its contribution
depends on the shape of the sample and in thin films the shape
anisotropy often results in the in-plane alignment of the moments.
The experimental evidence presented in this paper unambiguously
proves that perpendicular orientation of spontaneous magnetization
is realized in \emph{some} of the investigated (Ga,Mn)As/GaAs
layers. Importantly, similar studies\cite{Mune93} showed that PMA
is seen in \emph{most} (III,Mn)V layer grown such that they are
under tensile biaxial strain. As all are very thin layers
(typically a fraction of $\mu$m thick) and of macroscopic lateral
dimensions, such an experimental finding points to the existence
of a very strong microscopic mechanism that counteracts the shape
imposed in-plane arrangement of the magnetization.

Undoubtedly, the sign of the biaxial strain is one of the factors
that plays a role in determining of the direction of the magnetic
anisotropy. Indeed, at sufficiently high hole concentrations, this
is often the dominant factor. However, our experimental results
clearly show that in general, additional factor such as hole
concentration and temperature also play a role, and that the
anisotropy is determined by a combination of these factors. In
fact, we will argue that sensitivity to epitaxial strain, hole
density and temperature is an ubiquitous property of carrier
mediated ferromagnetism and is solely due to the anisotropy of the
carrier-mediated exchange interaction reflecting the anisotropic
properties of the top of the valence band. This should not be too
surprising given that we are dealing with magnetically diluted
systems, and that the shape anisotropy fields that must be
overcome are not particularly strong. In our case, as for thin
films, the shape anisotropy energy per unit volume is given by: $E
= \frac{1}{2}\mu_0 M_S^2 \cos^2\theta$ ($M_S$ is the saturation
magnetization and $\theta$ is the angle that $M_S$ subtends to the
plane normal), which gives the anisotropy field $\mu_0 H_A = \mu_0
M_S$ only of about 0.06~T for 5~\% (Ga,Mn)As, as compared to 2.2~T
for iron.

\begin{figure}
\includegraphics[width=3.4in]{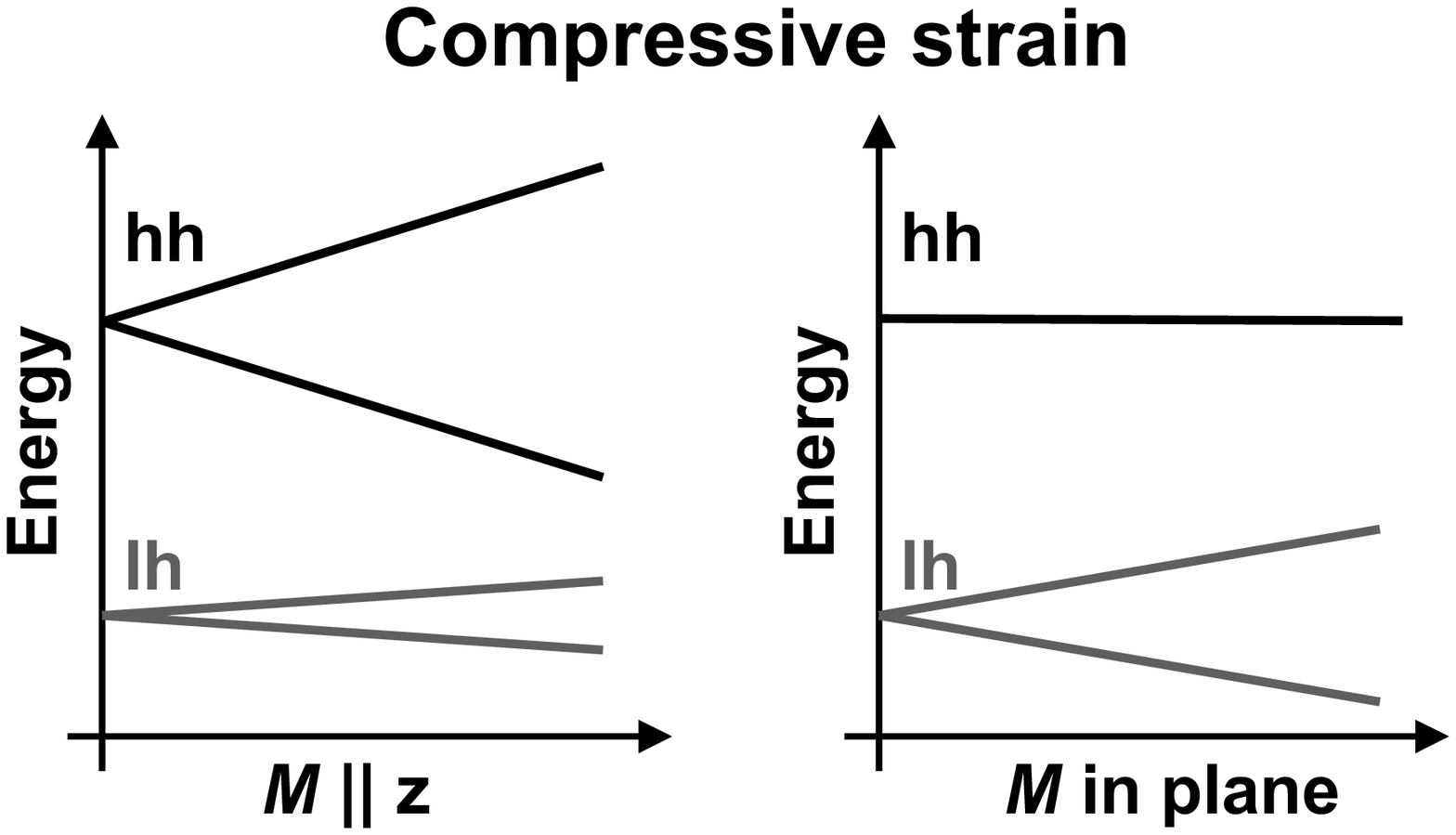}
\caption{\label{Pfig1}Scheme of valence band splitting in
tetrahedrally coordinated semiconductors for compressive strain
and for two orientations of magnetization $M$ with respect to sample
plane.}
\end{figure}

Knowing the p-d exchange energy and $k\cdot p$ parameters of the
valence band it is possible to compute magnetic anisotropy energy
in the studied compounds.\cite{Diet00,Diet01a,Abol01} In fact the
published results agree with the experimental data with remarkable
good accuracy.\cite{Liu03,Boukari02,Moore03,VDorpe04}
Nevertheless, it is instructive to consider a simplified case of
the model that is the nearly empty top of the valence band in
biaxial strained zinc-blende compounds. When the strain is present
the valence band splits and the energetic distance between the
heavy-hole $j_z = \pm 3/2$ and light-hole $j_z =\pm 1/2$ subbands
depends on strain, see Fig.~1. For the biaxial compressive strain
the ground state subband assumes a heavy-hole character. Then, if
only the ground state subband is occupied, the hole spins are
oriented along the growth direction. Now, since the p-d exchange
interaction has a scalar form, $H_{pd} \sim s\cdot S$, the
in-plane Mn spin magnetization $M$ will not affect the heavy-hole
subband. This means that perpendicular magnetic anisotropy is
expected, since only for such magnetization orientation can the
holes lower their energy by the coupling to the Mn spins. In the
opposite, tensile strain case, the in-plane component of the hole
spin is greater than the perpendicular component, so a stronger
exchange splitting occurs for the in-plane orientation of $M$.
Hence, if only the light-hole subband is occupied -- the in-plane
anisotropy is expected. It is worth remarking here, that PMA is
not a unique property of (III,Mn)V ferro-DMS. PMA is typically
realized in compressive strained (Cd,Mn)Te
QW\cite{Diet97,Boukari02} and the recent work showed that in
(II,Mn)VI/II,VI structures the magnetic anisotropy can be tailored
by adequate strain engineering.\cite{Kossa04} For the compressive
strain a ferromagnetic state related splitting of the luminescence
line is found only for the perpendicular orientation. However,
when large enough tensile strain was built into that system the
in-plane direction of the easy axis has been observed.

By nature (III,Mn)V DMS systems are heavily populated with holes.
But even when the Fermi energy is comparable or larger than the
heavy hole - light hole splitting, that is when the strong mixing
takes place, we can still talk about either heavy- or light hole
like character of the holes. Therefore the lines of reasoning
sketched above remain valid to a large extend in (III,Mn)V and
such a simplified approach can still serve as a helpful guideline.
So, we are going to use it again to sketch how in (III,Mn)V
ferro--DMS the direction of the magnetic easy axis can be set or
altered  by changing hole density and/or temperature.

\section{Reorientation of magnetic easy axis}

The Zener model dictates that Mn moments, or more accurately their
collective macroscopic magnetization, adjusts its orientation to
minimize the total energy of the carriers required to support
ferromagnetic ordering of the Mn ions. In particular, depending on
the Fermi level position within the valence band and/or the value
of the exchange splitting (that is depending on magnetization and
thus also on temperature) different orientations of magnetization
can be required to drive the system to its energy minimum.
Therefore, by changing hole concentration or temperature, the
corresponding changes of the overall orbital momentum of the hole
liquid may force a \emph{spontaneous} reorientation of
magnetization. It is relatively easy to trace such an effect when
we consider hole-concentration-induced reorientation. Using again
Fig.~1 we note that by introducing more holes into the system we
populate the second, light-hole, subband for which the
ferromagnetic state can only be realized with the in-plane
magnetization orientation. This effect gets even stronger since
the heavy holes acquire a light hole character on increasing Fermi
energy $E_{\mbox{\tiny F}}$, and it is therefore expected that at
some 'critical' value of the hole concentration, it will be more
favourable for the holes to experience  in-plane Mn magnetization.
So, an isothermal change of the hole density\cite{Ohno00} can lead
to the out-of-plane to in-plane reorientation of the magnetic
anisotropy, providing the population of light-holes gets large
enough. A similar mechanism operates when the temperature is used
as a handle for the easy axis switching. The only difference is
that this time the hh/lh population is changed via temperature
induced changes of the valence subbands exchange splitting. Now,
since the spin-splitting is proportional to Mn magnetization
$M_{\mbox{\tiny S}}(T)$ that varies according to the
Brillouin-like function, the character of magnetic anisotropy
depends on the temperature. Accordingly, in compressively strained
structures  PMA occurs at low both temperatures and hole
concentrations, while otherwise IMA will be realized. This implies
that there exists a class of samples for which the material
parameters are such that within the experimental temperature range
a reorientation of the easy axis from easy z-axis to easy plane
should occur on increasing temperature. An obvious question arises
then about the expected magnitude of changes of hole concentration
and temperature required for the reorientation to take place. At
this point, however, we want to strongly underline that despite
our general understanding of the physical mechanisms responsible
for the studied effects, the strong mixing of the valence band
states results in such a large anisotropic and non-parabolic
valence subbands dispersions that does not allow us to specify a
single hh/lh ratio and/or hole density/magnetization/temperature
values necessary to trigger magnetization easy axis switch. This
has to be computed up to the fullest possible extent of the mean
field model and for each and particular sample individually.

\section{Results and discussion}

Figures 2(a) and 2(b) present typical hysteresis loops for the
non-annealed (Ga,Mn)As sample which exhibits the temperature
driven change from PMA to IMA. First, let us note that the sample
does exhibit a perpendicular easy axis at low temperatures. As
Fig.~2(a) presents, a perfect square hysteresis is obtained when
the magnetic field is perpendicular to the film surface and an
elongated loop is seen when the in-plane orientation is probed.
Remarkably, a reverse behavior is observed \emph{for the same
sample} at higher temperatures (Fig.~2b). These findings
demonstrate that the easy axis flips from the perpendicular to
in-plane orientation on increasing temperature.

\begin{figure}
\includegraphics[width=3.4in]{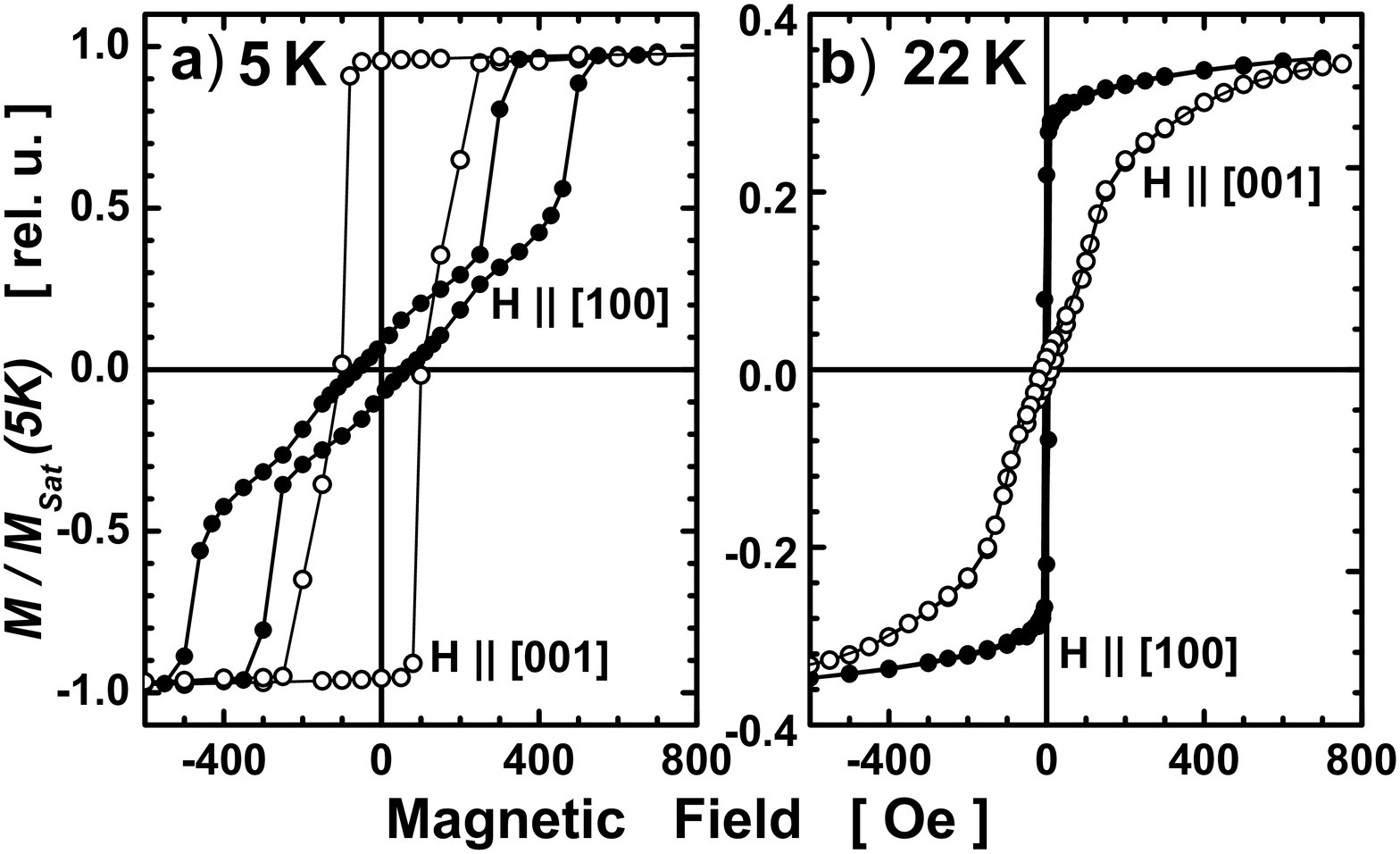}
\caption{\label{fig1} Field dependence of magnetization normalized
by its saturation value at 5 K for two orientations of the film in
respect to the magnetic field at two temperatures (a,b) for as
grown Ga$_{0.947}$Mn$_{0.053}$As. Full and empty dots denote the
data taken for the magnetic field along the [100] and [001]
crystal direction, respectively. Note the flip of the easy axis
direction from the perpendicular to the in-plane orientation on
increasing temperature.}
\end{figure}

\begin{figure}
\includegraphics[width=3.in]{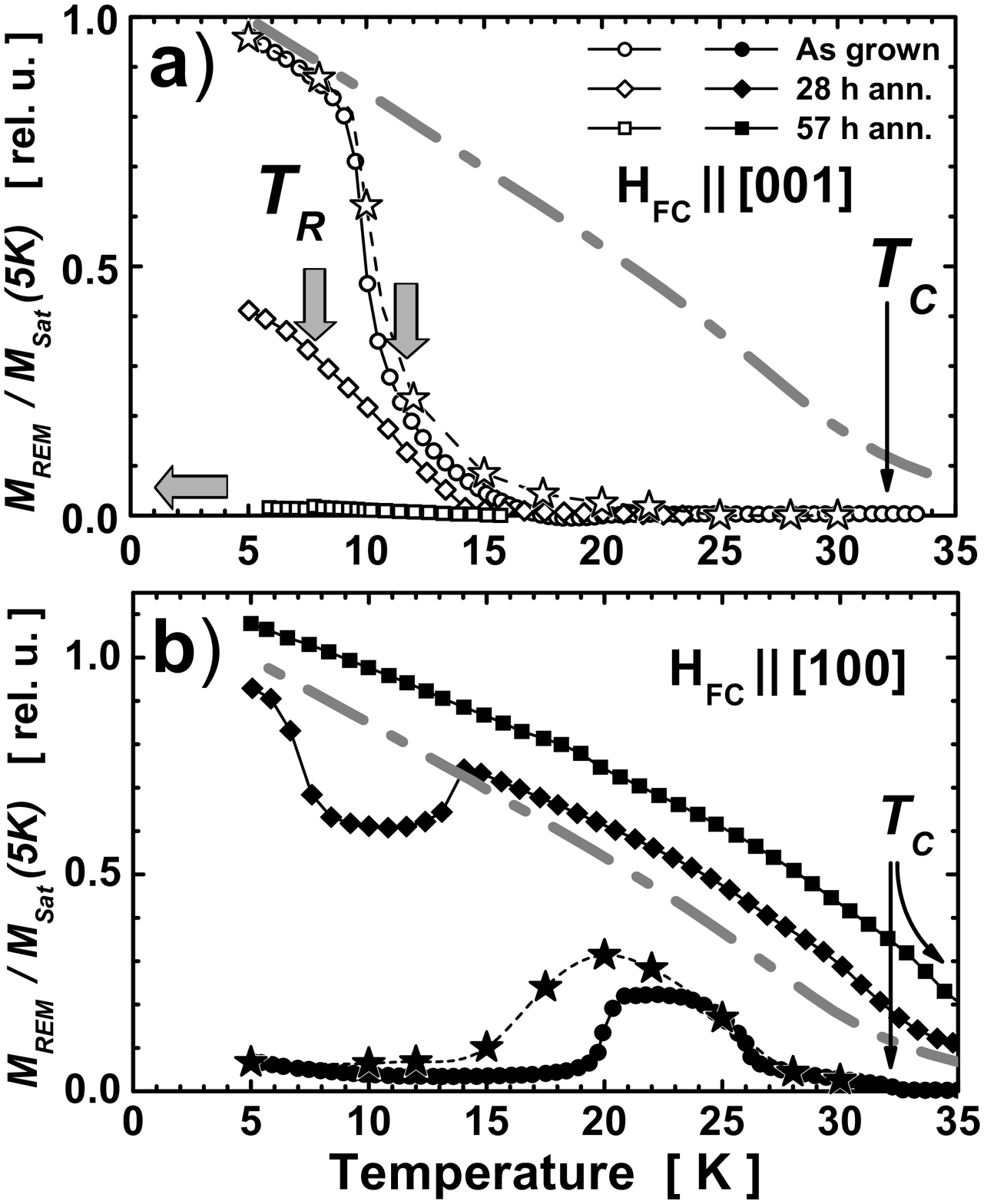}
\caption{\label{xfig2a} Temperature dependence of the remanent
magnetization as measured in perpendicular [001] (a) and in plane
[100] (b) configuration for the Ga$_{0.947}$Mn$_{0.053}$As sample
prior to annealing (circles) and after annealing (diamonds and
squares). Two experimental methods are presented. Temperature
dependent $M_{\mbox{\tiny{REM}}}$ (circles), the sample is cooled
down through $T_{\mbox{\tiny C}}$ in the field ${\bm
H}_{FC}=1$~kOe, which is at least by a factor of ten higher than
the coercive field $H_{\mbox{\tiny{C}}}$. Then, the field is
removed at 5~K, and the measurement of the magnetization component
$M_{\mbox{\tiny {REM}}}$ along the direction of ${\bm
H}_{\mbox{\tiny{FC}}}$ commences on increasing temperature in the
residual field ${\bm H}_{\mbox{\tiny {r}}}\leq 100$~mOe.
Alternatively, (Isothermal $M_{\mbox{\tiny{REM}}}$, stars) the
same $M_{\mbox{\tiny {S}}}$ component at selected temperatures
$T<T_{\mbox{\tiny C}}$ is obtained by field cooling from above
$T_{\mbox{\tiny C}}$ to $T$ and then removing the external
magnetic field. Both methods are seen to give essentially the same
results. Note that upon annealing the development of the in-plane
component of $M_{\mbox{\tiny{REM}}}$ is accompanied by an
equivalent quench of the perpendicular one. Bulk arrows mark the
reorientation temperature $T_{\mbox {\tiny R}}$ when the
cross-over to in-plane magnetic anisotropy takes place. The thick
dashed grey lines mark result of FC measurements at 1~kOe for the
as grown layer, and thus mimic the temperature dependence of the
saturation magnetization.}
\end{figure}

In order to trace the reorientation of spontaneous magnetization
$M_{\mbox{\tiny{s}}}$ more closely, we have examined the
temperature dependence of remanent magnetization $M_{\mbox{\tiny
{REM}}}$, measured along selected crystallographic directions
according to the procedures described in the caption to Fig.~3. As
evidenced in the figure we find $M_{\mbox{\tiny{REM}}}$ at low
temperatures to be tenfold larger for the perpendicular
experimental arrangement, ${\bm H}_{\mbox{\tiny{FC}}}\parallel
[001]$, compared to the case of the parallel one, ${\bm
H}_{\mbox{\tiny {FC}}}\parallel [100]$. This reconfirms the
appearance of PMA at low temperatures in this film despite the
presence of a sizable compressive strain. Large values of
perpendicular $M_{\mbox{\tiny{REM}}}$ hold until a certain
temperature ($\cong 10$~K in this case) at which a sudden drop of
the signal is detected. This is followed by a slow decay, which we
attribute to a material inhomogeneity. Further warming does not
change this situation, perpendicular $M_{\mbox{\tiny{REM}}}$ stays
vanishingly small until $T_{\mbox{\tiny C}}$. Clearly,
$M_{\mbox{\tiny{s}}}$ rotates out of [001] direction, and locks
itself in the plane of the layer somewhere within this temperature
range. We name the temperature at which this rotation takes place
the reorientation temperature ${\bm T}_{\mbox{\tiny {R}}}$.
However, the vanishing of perpendicular $M_{\mbox{\tiny{REM}}}$ is
not accompanied by an immediate development of parallel
$M_{\mbox{\tiny{REM}}}$. This does not contradict our
understanding, since due to equivalence of all in-plane easy cubic
directions, the flip of magnetization into the plane results in a
{\em demagnetized} state, characterized by the presence of closure
domains, and thus by small overall sample $M_{\mbox{\tiny{REM}}}$.
Actually, a sizable increase of the parallel $M_{\mbox{\tiny
{REM}}}$ between $T_{\mbox {\tiny R}}$ and $T_{\mbox{\tiny C}}$
reflects only the presence of the residual field ${\bm
H}_{\mbox{\tiny{r}}}$ generated in our magnet. This small field,
by breaking up the closure domains, starts to magnetize the film
at temperatures when the domain wall propagation mechanisms get
effective. However, since ${\bm H}_{\mbox{\tiny{r}}}$ assumes
extremely small value in our system, the sample gets only
partially magnetized, which accounts for the differences observed
at this temperature range between temperature-dependent REM and
isothermal REM measurements, as seen in Fig.~3(b).

As already noted, the temperature-induced cross-over from PMA to
IMA described above has been found in  other samples of
(Al,Ga,Mn)As and (Ga,Mn)As.\cite{Taka02,Sawi03} Remarkably, the
opposite behavior occurs in (In,Mn)As under a tensile strain,
where the easy axis flips from in-plane to out of plane on
warming.\cite{Endo01,Liu04a} As sketched in the previous section,
this stems from the fact that the direction of orbital momentum,
which controls magnetic anisotropy, differs for the heavy and
light holes, whose relative concentration depends not only on the
sign of biaxial strain but on the magnitude of the spin splitting
as well. Importantly, striking temperature dependent anisotropy in
these systems can, in fact, be inferred from numerical
computations presented in the previous theoretical
work.\cite{Diet01a} In particular, according to Fig.~10 of
Ref.~\onlinecite{Diet01a}, there is a range of the hole
concentrations for which the easy axis of compressed films is
expected to switch from the perpendicular to the in-plane
direction on decreasing the valence band spin-splitting (that is
on increasing temperature, while the switching in the opposite
sense is predicted for the tensile strain.

Figure 4 presents the computed reorientation temperature $T_{\mbox
{\tiny R}}$ as a function of the hole concentration $p$ for the
two studied samples. The theoretical model and material parameters
described in detail previously\cite{Diet01a} are adopted to obtain
this phase diagram. Shape anisotropy is taken into account, and it
shifts the PMA $\rightarrow$ IMA phase boundary by about 20~\%
towards lower $p$ values. The same theory is employed to determine
the magnitude of $p$ from the experimental values of $x$ and
$T_{\mbox{\tiny C}}$. We see that the theoretical model confirms
the appearance of PMA and describes correctly temperature driven
phase transformation PMA $\rightarrow$ IMA in the case of the $x
=5.3$\% sample prior to annealing. According to theory, such a
phenomenon should vanish for higher values of $p$. Indeed, as seen
from Fig.~3, if the annealing time and thus the magnitude of
$T_{\mbox{\tiny C}}$ and $p$ are sufficiently large, these samples
exhibit only IMA.

At this point it is worth to emphasize that although the
reorientation transition is the general feature of heavily doped
ferro-DMS, it is only a sample specific property. In particular,
for a given strain if the hole concentration is either too small
or too large no reorientation transition is expected for any value
of magnetization (temperature). In fact, as Fig.~4 demonstrates,
the range of hole densities for which the reorientation can
occur is quite narrow. On the other hand, for an appropriate
combination of strain and hole concentration, even a minute change of
magnetization (temperature) switches the easy axis between the two
directions. This feature is confirmed by a comparison of our
$M_{\mbox{\tiny{REM}}}(T)$ and quasi-$M_{\mbox{\tiny S}}(T)$ data
for the as-grown sample. As seen in Fig.~3, $M_{\mbox{\tiny S}}$
is undoubtedly a smooth and slowly varying function of temperature,
and despite of this the reorientation does take place.

\begin{figure}
\includegraphics[width=3.3in]{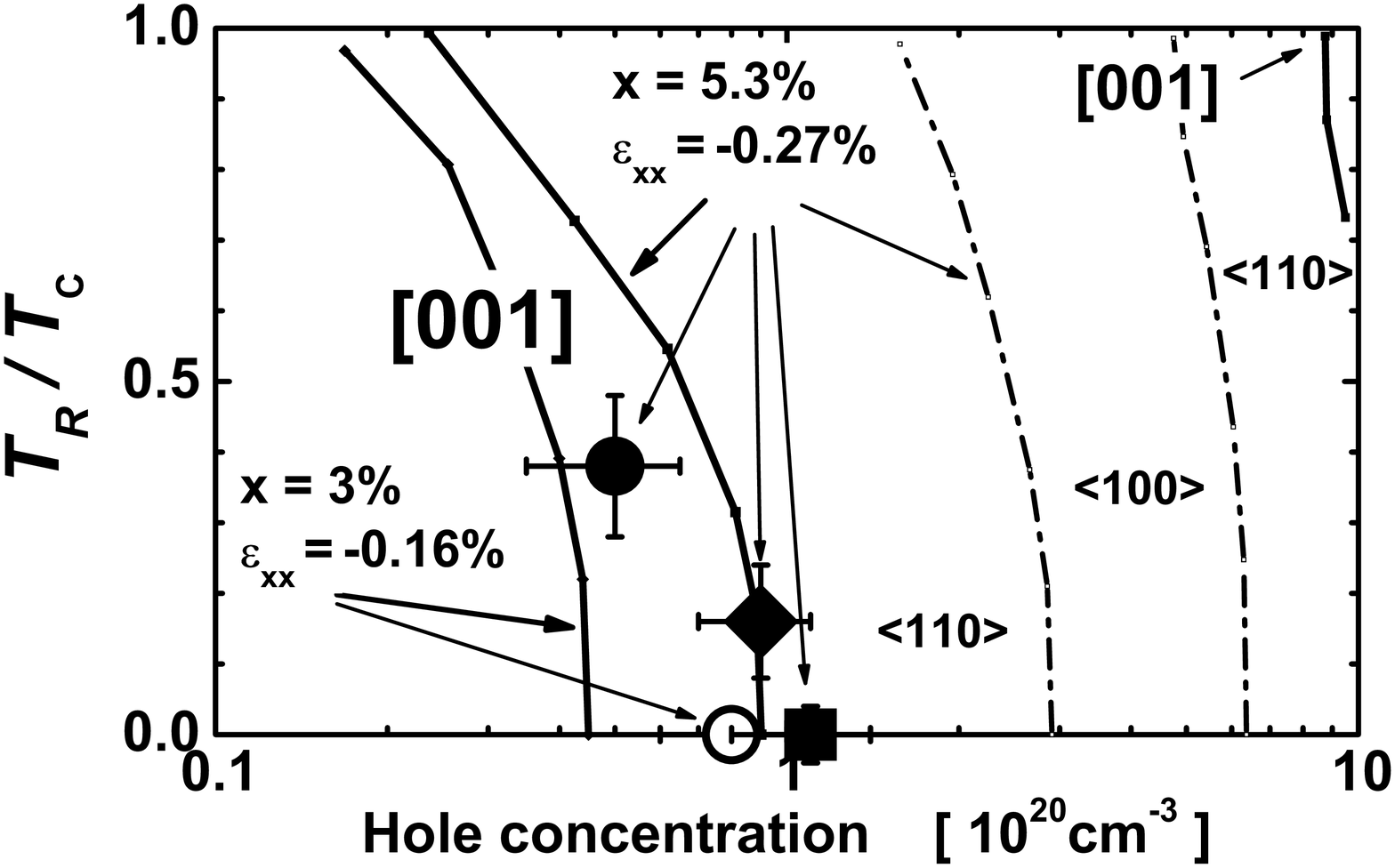}
\caption{Experimental (full points, taken from Fig.~3 for $x
=5.3$~\% sample) and computed values (thick lines) for the ratio
of reorientation and Curie temperatures in Ga$_{1-x}$Mn$_x$As for
perpendicular to in-plane magnetic anisotropy transition. For $x
=3$~\% sample (the open circle) this transition is not detected
above 5~K, in an agreement with the presented calculations. Dashed
lines mark expected temperatures for the in-plane reorientation of
the easy axis between $\langle 100\rangle$ and $\langle
110\rangle$ directions.}
\end{figure}

According to the discussion above, the easy axis assumes the
in-plane orientation for typical carrier concentrations in
(Ga,Mn)As/GaAs. In this case, according to the theoretical
predictions presented in Fig.~4 as well in Fig.~9 of
Ref.~\onlinecite{Diet01a} and in Fig.~6 of
Ref.~\onlinecite{Abol01}, the fourfold magnetic symmetry with the
easy axis is expected to switch between $\langle 100\rangle$ and
$\langle110\rangle$ in-plane cubic axes as a function of $p$ or
$T$. This biaxial magnetic symmetry is indeed observed at low
temperatures, however with the easy axis assuming exclusively
$\langle100\rangle$ in-plane orientation, as observed by us and
others.\cite{Liu03,Kats98,Hrab02,Welp03,Moore03,Welp04} To our
best knowledge, no
$\langle100\rangle\Leftrightarrow\langle110\rangle$ transition has
been detected to date. This may indicate that the anisotropy of
the hole magnetic moment, neglected in the theoretical
calculations,\cite{Diet01a,Abol01} stabilizes the
$\langle100\rangle$ orientation of the easy axis. It is also
possible that the stabilization energy comes from broken magnetic
bonds at the film surfaces, an effect put into evidence in the
case of (Fe,Co)/GaAs films.\cite{Dumm02} However, whether any of
these models explains simultaneously the reported recently
$\langle110\rangle$ biaxial symmetry in (In,Mn)As/(In,Al)As
films,\cite{Liu04} remains to be shown.

\begin{figure}
\includegraphics[width=3.3in]{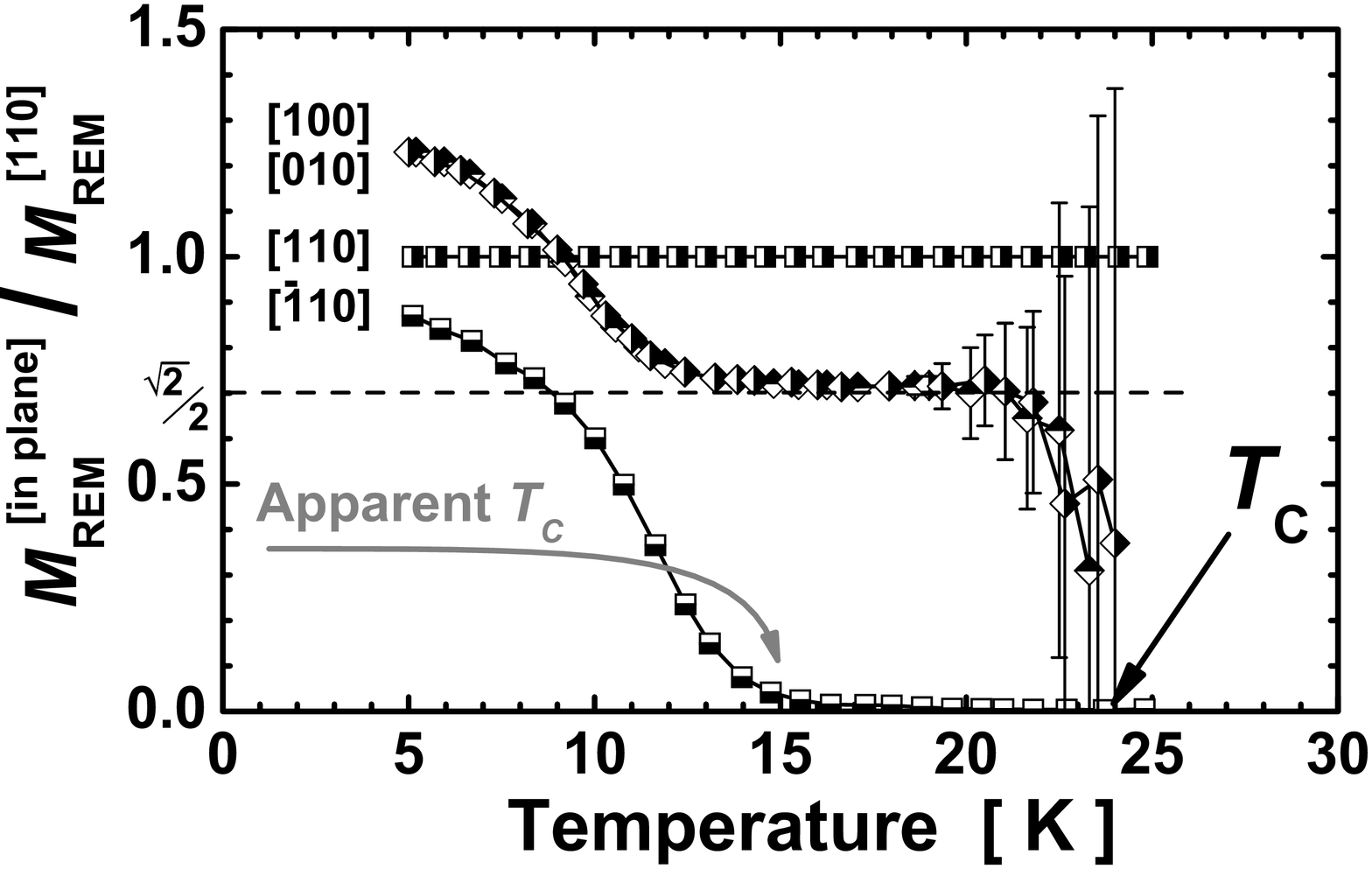}
\caption{\label{fig4}Experimental evidence for the uniaxial
anisotropy along [110] direction in Ga$_{0.97}$Mn$_{0.03}$As film.
The magnetic remanence is measured for four major in-plane
directions and its magnitude is normalized by the data of the
[110] case. Note, that the sudden drop of $M_{\mbox{\tiny {REM}}}$
along $[\bar{1}10]$ at $T \ll {\bm T}_{\mbox{\tiny {C}}}$ may
wrongly indicate too low value of ${\bm T}_{\mbox{\tiny {C}}}$, if
only this orientation is probed.}
\end{figure}
Our magnetization data reveal also the existence of a uniaxial
anisotropy in films of (Ga,Mn)As/GaAs. As shown in Fig.~4,
$M_{\mbox{\tiny {REM}}}$ measured along the $[\bar{1}10]$
direction vanishes completely above 15~K indicating that this is
the hard direction in this film. We also note that when
$M^{[\bar{1}10]}_{\mbox{\tiny {REM}}}$ vanishes the
$M^{[100]}_{\mbox{\tiny {REM}}}/M^{[110]}_{\mbox{\tiny {REM}}}$
ratio drops to $\sqrt{2}/2$, the value expected for the easy axis
along $[110]$. Since the cubic-like anisotropy energy is
proportional to $M_{\mbox{\tiny {s}}}^4$ whereas the uniaxial one
to $M_{\mbox{\tiny {s}}}^2$, the latter dominates at high
temperatures, where $M_{\mbox{\tiny {s}}}$ is small. We note that,
due to the biaxial strain, the initial T$_{\mbox{\tiny{d}}}$ point
symmetry of zinc blende structure is lowered to
D$_{\mbox{\tiny{2d}}}$, for which {\em no} uniaxial IMA is
expected. Furthermore, the magnitude of the corresponding
anisotropy field appears to be independent of the film thickness,
for both as large as 7~$\mu$m\cite{Welp04} and as low as
25~nm\cite{Sawic04}, what, in particular, rules out the effect of
Mn oxide accumulated at the free surface.\cite{Edmo04,Furdy04}
Recently the Argone - Notre Dame team\cite{Welp03,Welp04} has
advocated for an effect connected with surface reconstruction
induced preferential Mn incorporation occurring at every step of
layer-by-layer growth. At the same time we point out that a
unidirectional character of the growth process\cite{Welp03} and/or
differences between (Ga,Mn)As/GaAs and (Ga,Mn)As/vacuum interfaces
may lower symmetry to C$_{\mbox{\tiny{2v}}}$ where the three
principal directions are: [001], [110] and $[\bar{1}10]$. Since in
C$_{\mbox{\tiny{2v}}}$ they are not equivalent,  the $[110]
\Leftrightarrow$ $[\bar{1}10]$\ symmetry gets broken, conforming
with the presented results.

In summary, our studies have demonstrated the rich characteristics
of magnetic anisotropies in (Ga,Mn)As/GaAs, which---in addition to
epitaxial strain---vary with the hole and Mn concentrations as
well as with the temperature. According to
theory,\cite{Diet00,Diet01a,Abol01} these reflect spin anisotropy
of the valence band subbands whose shape varies with strain, while
the splitting and population depend on magnetization and hole
concentration. In particular, the temperature-driven reorientation
of the easy axis reflects the equipartition  of the valence
subband population at $T$ approaching $T_{\mbox{\tiny C}}$ and,
therefore, reconfirms the crucial role of the valence band holes
in the ferromagnetism of (III,Mn)V systems. At the same time, our
findings have provided the magnetic corroboration for the
existence of uniaxial in-plane anisotropy. By group theoretical
considerations this unexpected anisotropy can be linked to the
top/bottom symmetry breaking, an issue calling for a microscopic
modelling. Furthermore, the temperature-induced switching of the
easy axis direction revealed by our findings explains the origin
of non-standard temperature dependencies of measured
magnetization, and indicates that the meaningful determination of
$T_{\mbox{\tiny C}}$ requires measurements for various crystal
orientations.  We note also that since the direction of
spontaneous magnetization depends on the hole concentration which,
in turn, can be varied by the electric field\cite{Ohno00} or light
irradiation,\cite{Liu04a} it appears possible to reverse the
magnetization in a field-effect transistor having a ferromagnetic
semiconductor channel.

\begin{acknowledgments}
The authors thank J. Ferr\'e, H. Ohno, and W. Van Roy  for
valuable discussions. Support of the FENIKS project
(EC:G5RD-CT-2001-0535) is gratefully acknowledged.

\end{acknowledgments}


\end{document}